%% file: tiling.tex
\documentclass{article}

\usepackage{amssymb,amsmath,amsfonts,graphicx,latexsym,wrapfig,fullpage}

\input{macros.tex}

\newtheorem{lemma}{Lemma}
\newtheorem{theorem}{Theorem}

\begin{document}

\title{Tile-Packing Tomography is {\NP}-hard}

\author{Marek Chrobak
\and 
Christoph D\"urr%
\thanks{CNRS and Lab. of Computer Science of the \'Ecole Polytechnique,
France}
\and
Flavio Gu\'{\i}\~nez%
\thanks{Operations and Logistics Division, Sauder School of Business,
  Canada}
 \and 
Antoni Lozano%
\thanks{Logic and Programming Research Group, UPC. Barcelona Tech,
  Catalonia, Spain}
 \and 
Nguyen Kim Thang%
\thanks{LAMSADE, Universit\'e Paris Dauphine, France}}

\maketitle

\begin{abstract}
Discrete tomography deals with reconstructing finite spatial objects
from their projections. The objects we study in this paper are
called tilings or tile-packings, and they consist of a number of disjoint copies 
of a fixed tile, where a tile is defined as a connected set of grid
points. A row projection specifies how many grid points are covered by
tiles in a given row; column projections are defined analogously.
For a fixed tile, is it possible to reconstruct its tilings from their
projections in polynomial time? It is known that the answer to this
question is affirmative if the tile is a bar (its width or height is $1$),
while for some other types of tiles {\NP}-hardness results have been shown
in the literature. In this paper we present a complete solution to this
question by showing that the problem remains {\NP}-hard for \emph{all} 
tiles other than bars.

\end{abstract}


\section{Introduction}

Discrete tomography deals with reconstructing finite spatial objects
from their low-dimensional projections. Inverse problems of this nature arise
naturally in medical computerized tomography,
electron tomography, non-destructive quality control, timetable
design and a number of other areas.
This wide range of applications inspired significant theoretical
interest in this topic and led to studies of computational
complexity of various discrete tomography problems. For an extensive
and detailed coverage of practical and theoretical aspects of
this area,  we refer readers to the book by Kuba and Herman, see
\cite{Kuba.Herman:discr_tomo1999,Herman.Kuba:discr_tomo2007}.

In this paper we consider the problem of reconstructing a tile packing
from its row and column projections.  Formally, consider the integer
grid of dimension $m\times n$, consisting of all cells $(i,
j)\in[0,m)\times[0,n)$. 
Every cell $(i, j)$ is adjacent to its
neighbor cells 
\[
  \begin{array}{ccc}
&(i - 1, j),
\\
 (i, j - 1),&& (i, j + 1),\\
&(i + 1, j),
\end{array}
\]
whichever of those are present in the grid.  Alternatively, one can
think of each $(i,j)$ as a cell in an $m\times n$ matrix. 
In the
paper, we will often use the matrix notation and terminology, using
terms ``row'' and ``column'', with rows numbered top-down and columns
numbered from left to right, so that the upper-left cell is $(0,0)$.

We define a \emph{tile} to be any finite connected set $T$
of grid cells.  By ``connected'' we mean that for any two cells of $T$
there is a path inside $T$ between these cells, where any two
consecutive cells on this path are adjacent. 
The \emph{width} and
\emph{height} of $T$ are defined in the obvious manner, as the
dimensions of the smallest $h\times w$ rectangle containing $T$.  If
$w=1$ or $h=1$, then $T$ is called a \emph{bar}.  By 
\[
   T+({i},{j}) = \braced{(x+i,y+j)\suchthat (x,y)\in T}
\] 
we denote the translation of
$T$ by the vector $({i},{j})$. $T+(i,j)$ is called a
\emph{(translated) copy} of $T$, and $(i,j)$ is the \emph{position} of 
this copy. Later in the paper, we will sometimes refer to
$T+(i,j)$ as a ``tile" -- which is somewhat ambiguous but more
intuitive than a ``copy"

\begin{figure}[ht]
\begin{center}
\includegraphics[width=3.5cm]{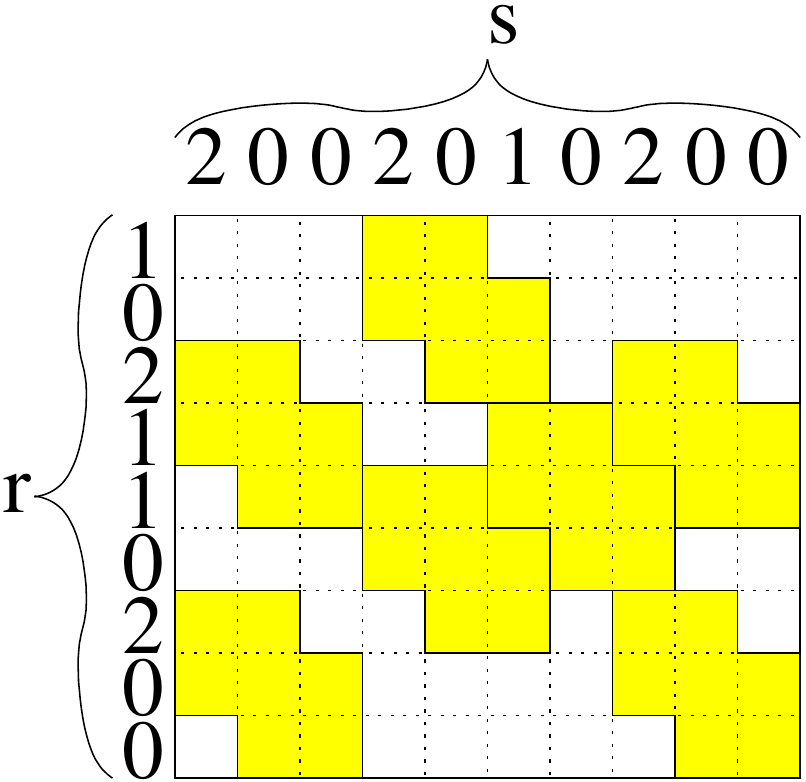}
\end{center}
\caption{A tile packing of the $9\times 10$ grid and its
  projections. By convention the leftmost upper cell is $(0,0)$. }
\label{fig:exampleProj}
\end{figure}

A \emph{tile packing of the $m\times n$ grid using $T$} --- or
a \emph{$T$-packing}, in short, if $m$ and $n$ are understood from
context --- is a disjoint partial covering of the grid with translated
copies of $T$.  Formally, a $T$-packing is defined by a set $D$ of
translation vectors such that all translated copies $T+({i},{j})$, for
all $({i},{j})\in D$, are contained in the $m\times n$ grid and are
pairwise disjoint. 
We do not require the tiles to completely cover the grid. We stress here that
what we call a $T$-packing here has been
sometimes called a \emph{partial tiling} in the literature;
see, for example, \cite{FrosiniSimi_05}.
An example of a $T$-packing is shown in Figure~\ref{fig:exampleProj}.
Without loss of generality,
throughout the paper, we will be assuming that the tile $T$ used in
packing is in a \emph{canonical position} in the upper-left corner of
the grid, that is $\min\braced{x\suchthat (x,y)\in T} =
\min\braced{y\suchthat (x,y)\in T} = 0$.

To simplify notation, instead of counting 
how many grid points are covered by
tiles in a given row (or column), we count how many tiles \emph{start} in a given
row (column), which is equivalent up to some base-change.
So the \emph{row} and \emph{column projections} of a packing $D$
are defined as a pair $r\in\mathbb{N}^{m}$ and $s\in\mathbb{N}^{n}$ 
of vectors such that 
\begin{eqnarray*}
	r_{i} \;=\; \barred{\braced{ j \suchthat (i,j)\in D}}
	\quad\textrm{and}\quad
	s_{j} \;=\; \barred{\braced{ i \suchthat (i,j)\in D}}.
\end{eqnarray*}
For example, consider tile 
$T = \braced{(0,0), (0,1), (1,0),(1,1), (1,2), (2,1), (2,2)}$.
Figure~\ref{fig:exampleProj} shows an example of a $T$-packing.
This packing is 
\[D = \braced{(0,3), (2,0), (2,7), (3,5), (4,3),(6,0), (6,7)}.\]

\smallskip

We study the problem of reconstructing tile packings from its
horizontal and vertical projections. More formally, for any fixed tile
$T$, the problem is defined as follows:

\begin{description}
\item[Tile Packing Tomography Problem ({\TPTP{T}}):] 
The instance of
{\TPTP{T}} consists of vectors $r\in {\mathbb N}^{m}, s\in {\mathbb N}^{n}$. The objective is to decide if there is a $T$-packing $D$ whose projections are $r$ and $s$.
\end{description}

This problem has been introduced in~\cite{CCDW01} and shown to be
{\NP}-hard for some particular
tiles. In~\cite{DurrGuinez:Reconstructing-3-Colored}, the proof
technique has been adapted to show {\NP}-hardness for any rectangular
tile, i.e. a tile that consists of all cells
$(i,j)\in[0,h)\times[0,w)$ for some dimensions $w,h\geq 2$.

On the positive side, the classical work of
Ryser~\cite{Ryser63} on projections of 0-1 matrices provides a
characterization of vectors that correspond to projections of
$T$-packings for the special case when $T$ is a single cell,
and provides a simple polynomial-time algorithm for that
case. The ideas from \cite{Ryser63} were extended in
\cite{durr.goles.ea:03:tiling*,Picouleau99} to the case when $T$ is a
bar. In~\cite{Brunetti.Costa.ea:binary-adjacency}, polynomial-time
algorithms were given for restricted special cases. The complexity
status was unknown for all other tiles, and the current paper completes
the picture by proving the following theorem.


\begin{theorem}\label{thm: np-completeness}
Problem~{\TPTP{T}} is $\NP$-complete for any tile $T$ that is not a bar.
\end{theorem}

The general structure of our proof resembles those introduced in
\cite{CCDW01} and \cite{DurrGuinez:Reconstructing-3-Colored}, although
the reductions we present are substantially more difficult, since the
generality of our result means that we cannot take advantage of a
specific shape of the tile, and that we need to base the construction
on some generic properties shared by infinitely many tiles.  Our 
techniques take advantage of Ryser's structure results for 0-1
matrices, in combination with some arguments based on linear algebra.

After reviewing some background information in Section~\ref{sec: main
  tools}, we introduce the main idea of the reduction in
Section~\ref{sec:scheme}, by formulating the overall framework of the
reduction and conditions on $T$ required for this reduction to be
correct.  Then, in Section~\ref{sec: construction of block packings}
we show that all non-bar tiles satisfy these conditions.


\section{Main Tools}   \label{sec: main tools}

In this section we briefly review two concepts that will play a crucial
role in our proofs: affine independence and Ryser's theorem.


\paragraph{Affine independence.}
Vectors $v_1,v_2,...,v_k\in\reals^n$ are called \emph{affinely
  independent} if the unique solution of equations 
\[
\sum_{i=1}^k
\alpha_i = 0 \mbox{\hspace*{2em}and\hspace*{2em}}\sum_{i=1}^k \alpha_i v_i = 0
\]
 is $\alpha_1 =
\alpha_2 = ... = \alpha_k = 0$. It is easy to show that the following
three conditions are equivalent:
\begin{description}
	\item{(ai1)}  $v_1,v_2,...,v_k$ are affinely independent,
	\item{(ai2)} $v_2-v_1,v_3-v_1,...,v_k - v_1$ 	are linearly independent,
	\item{(ai3)} $(v_1,1), (v_2,1), ..., (v_k,1)$ are linearly independent.
\end{description}
We will refer to vectors $v_i-v_1$, $i=2,3,...,k$, in (ai2), as
\emph{difference vectors}. Condition (ai2) is useful in verifying
affine independence.  For example, $(1,1), (3,4), (5,5)$ are affinely
independent because the difference vectors $(3,4) - (1,1) = (2,3)$
and $(5,5)-(1,1) = (4,4)$ are linearly independent.

Condition (ai3) implies that if $v_1,v_2,...,v_k$ are affinely
independent then for any vector $v$ and constant $\beta$, the equations
\[
 \sum_{i=1}^k\alpha_i v_i = v\mbox{\hspace*{2em}and\hspace*{2em}}\sum_{i=1}^k \alpha_i = \beta
\]
 have
a unique solution $\alpha_1,\alpha_2,...,\alpha_k$.

For any vector $v\in \reals^n$ and any set of indices
$i_1,i_2,...,i_b\in[0,n)$, define the \emph{
  $(i_1,i_2,...,i_b)$-restriction of $v$} to be the vector
$v'\in\reals^b$ that consists only of the coordinates $i_t$,
$t=1,...,b$, of $v$. For example, the $(0,3,4)$-restriction of $v =
(4,3,1,0,7,9,5)$ is $v' = (4,0,7)$.  For any set of vectors
$v_1,v_2,...,v_k\in\reals^n$, to show that they are affinely
independent it is sufficient to show that their
$(i_1,i_2,...,i_b)$-restrictions are affinely independent, for some
set of indices $i_1,i_2,...,i_b$.


\paragraph{Ryser's theorem.}
Ryser \cite{ryser60:_matric_zeros_ones} studied the structure of 0-1
matrices with given projections. We adapt his characterization of
these matrices and express it in terms of tile packings.

Fix a tile $T$ and let $I\subseteq[0,m)$ be a set of rows and
$J\subseteq[0,n)$ a set of columns.  We say that a tile copy $T+(i,j)$
\emph{belongs to} $I\times J$ if $i\in I, j\in J$.  Note that here we
do not require inclusion of $T+(i,j)$ in $I\times J$.

Define $\xi_{I,J} =\max_D |D\cap (I\times J)|$, where the maximum is
taken over all $T$-packings $D$ of the $m\times n$ grid. Thus
$\xi_{I,J}$ is the maximum number of copies of $T$ that can belong to
$I\times J$ in a $T$-packing without overlapping (and without any
restriction on their projections).

For a set $I$ of rows, denote $r(I) = \sum_{i\in I}r_{i}$. Analogously, let
$s(J) = \sum_{j\in J}s_{j}$, for a set $J$ of columns. By 
$\barI = [0,m)-I$ and $\barJ = [0,n) - J$ we denote the complements of 
of sets $I$ and set $J$, respectively.

Consider a $T$-packing $D$ with projections $r,s$. Then we have
\begin{eqnarray*}
	r(I) - s(\barJ) 
		&=&
		|D\cap (I\times J)| - |D\cap (\barI\times\barJ)|.
\end{eqnarray*}
By definition, $|D\cap (I\times J)|\le \xi_{I,J}$. Therefore we obtain
the following lemma (inspired by \cite{ryser60:_matric_zeros_ones}).


\begin{lemma} \label{lem:ryser-adapted} Let $I$ be a set of rows and
  $J$ a set of columns.  If $r(I)-s(\barJ)=\xi_{I,J}$ then every
  $T$-packing $D$ with projections $r,s$ satisfies $|D\cap (I\times J)|
  = \xi_{I,J}$ and $|D\cap (\barI\times \barJ)| = 0$.
\end{lemma}	


\section{General Proof Structure}   \label{sec:scheme}

For each non-bar tile $T$, we show a polynomial-time
reduction from the \emph{3-Color Tomography Problem} introduced in
\cite{GaGrPr00} and shown to be {\NP}-hard in
\cite{DurrGuinez:Reconstructing-3-Colored}. In that problem, an object
to be reconstructed is a set of ``atoms'' (in our terminology, single
cells) colored red (R), green (G) or blue (B). The instance contains
separate projections for each color. The formal definition is this:

\begin{description}
\item[3-Color Tomography Problem ({\threeCTP}):] The instance consists
  of six vectors $r^R,$ $r^G,r^B \in {\mathbb N}^{m}$, $s^R,s^G,s^B\in
  {\mathbb N}^{n}$. The objective is to decide whether there is an
  $m\times n$ matrix $M$ with values from $\braced{R,G,B}$ such that,
  for each color $c\in\braced{R,G,B}$, $r^{c}_{x} = \barred{\braced{y
      \suchthat M_{xy}=c}}$ for each $x$ and $s^{c}_{y} =
  \barred{\braced{x \suchthat M_{xy}=c}}$ for each $y$.
\end{description}

From now on, assume that $T$ is some non-bar fixed tile of width $w$
and height $h$.  Let $\calI$ be an instance of {\threeCTP} for some
$m\times n$ matrix specified by six projections
$r^R,r^G,r^B,s^R,s^G,s^B$. We will map $\calI$ into an instance
$\calJ$ of {\TPTP{T}} for an $m'\times n'$ grid with projections $r$,
$s$, such that $\calI$ has a matrix $M$ with projections
$r^R,r^G,r^B,s^R,s^G,s^B$ if and only if $\calJ$ has a $T$-packing with projections $r,s$.

Without loss of generality we assume that for every color $c$ we
have $\sum_{x} r^{c}_{x}=\sum_{y} s^{c}_{y}$, for every row $x$ we have
$\sum_{c} r^{c}_{x}=m$, and for every column $y$ we have $\sum_{c}
s^{c}_{y}=n$.  Otherwise, $\calI$ is of course unfeasible, so we could
take $\calJ$ to be any fixed unfeasible instance of {\TPTP{T}}.

We now describe $\calJ$. We will choose a grid of size $m'\times n'$
for $m' = mk$ and $n' = n\ell$, where $k$ and $\ell$ are positive
integer constants to be specified later.  We will use the term
\emph{block} for a $k\times \ell$ grid. We can partition our $m'\times
n'$ grid into $mn$ rectangles of dimension $k\times \ell$, and we can
think of each such rectangle as a translated block. The rectangle
\(
  [xk,(x+1)k)\times [y\ell,(y+1)\ell)
\) 
will be referred to as block $(x,y)$.

Next, we need to specify the projections $r$ and $s$. We will describe
these projections in a somewhat unusual way, by fixing three packings
of a block denoted $D^R$, $D^G$, and $D^B$ (obviously, corresponding
to the three colors), and then expressing $r$ and $s$ as linear
combinations of these packings.  More specifically, denoting by
$\barr^c$ and $\bars^c$ the horizontal and vertical projections of
packing $D^c$, for each $c\in\braced{R,G,B}$, we define
\begin{align}
    r_{xk+i} \;=\; \sum_c r^c_{x} \cdot \barr^{c}_{i}
	\quad\textrm{and}\quad
    s_{y\ell+j} &= \sum_c s^c_{y} \cdot \bars^{c}_{j}  \label{def:proj-reduc}
\end{align}
for every $i\in[0,k)$, $j\in[0,\ell)$, $x\in[0,m)$, and
$y\in[0,n)$. The idea is that replacing each cell in a solution to
the {\threeCTP} instance $r^c,s^c$, for $c=R,G,B$, 
by its color-corresponding block gives a solution to the \TPTP{T} instance $r, s$.

To complete the description of the reduction, it still remains to
define the three packings $D^R$, $D^G$, and $D^B$. This will be done
in the next section. In the remainder of this section we establish
conditions that will guarantee correctness of our reduction.

Our three packings will be designed to satisfy the following two
requirements:
\begin{description}
	
\item{\emph{Requirement~1:}} Vectors $\barr^{R}, \barr^{G}, \barr^{B}$ are affinely 
independent and vectors $\bars^{R}, \bars^{G},$
  $\bars^{B}$ are affinely independent.  Note that, by property (ai3),
  this implies that for any vector $v$ there is at most one possible
  way to represent it in a form $v = n_R\barr^R+n_G\barr^G+n_B\barr^B$, 
where $n_R+n_G+n_B=n$. Naturally, an analogous statement holds for
column projections.

\item{\emph{Requirement~2:}} In any packing $D$ of $\calJ$ with
  projections $r,s$, the restriction of $D$ to each block of the grid
  has projections equal to $\barr^c, \bars^c$, for some
  $c\in\braced{R,G,B}$.

\end{description}


\begin{lemma}\label{lem:tiling-structure}
Assume that the three packings $D^R, D^G, D^B$ satisfy Requirements 1 and 2.
Then $\calI$ has a solution if and only if $\calJ$ has a solution. 
\end{lemma}

\begin{myproof}
$(\Rightarrow)$
Let $M\in\{R,G,B\}^{m\times n}$ be a solution to $\calI$.
We transform $M$ into the following packing $D$ for the $m'\times n'$ grid:
\begin{align*}
	D  = \bigcup_{x\in[0,m)} \bigcup_{y\in[0,n)}
						\parend{ D^{M_{xy}} + (xk,y\ell) }.
\end{align*}
In other words, if $M_{xy} = c$ then block $(x,y)$ of the $m'\times
n'$ grid contains a copy of $D^c$. By simple inspection, the
projections of $D$ are indeed equal to the vectors $r$ and $s$ in
(\ref{def:proj-reduc}).

$(\Leftarrow)$ For the converse, suppose that there is a packing $D$
with projections $r$, $s$.  By Requirement~2, every block of the
$m'\times n'$ grid has projections $\barr^c$ and $\bars^c$, for some
$c\in\braced{R,G,B}$. We then \emph{associate} this block with color
$c$.  We can thus define a matrix $M\in\braced{R,G,B}^{m\times n}$
such that $M_{xy}=c$ if block $(x,y)$ of $D$ is associated with color
$c$.

We now need to show that $M$ is a solution for $\calI$. To this end,
fix some arbitrary $0\leq x < m$ and consider vector 
\[
 v = (r_{xk}, r_{xk+1}, ..., r_{(x+1)k-1}),
\]
 which is the projection of the ``row''
of all blocks $(x,y)$, for all $y$. By the construction (\ref{def:proj-reduc}),
$v$ can be written as 
\[
          v = n_R \barr^{R} + n_G \barr^{G} +n_B \barr^{B},
\]
where $n_R = r^R_x$, $n_G=r^G_x$, and $n_B = r^B_x$.  Now, using
Requirement~1, we obtain that this representation is unique under the
assumption that $n_R+n_G+n_B=n$. We can thus conclude that the
projection of row $x$ of $M$ is correct, that is $|\braced{y\suchthat
  M_{xy} = c}| = r^c_x$ for all $c$.  By the same argument, column
projections of $M$ are correct as well, completing the proof.
\end{myproof}

In summary, to complete the proof for the given tile $T$, we need to
do this: (i) define a rectangular $k\times \ell$ block with three
packings $D^R, D^G, D^B$, (ii) show that the row projections of $D^R,
D^G, D^B$ and the corresponding column projections are affinely
independent (Requirement~1), and (iii) show that in any solution to
$\calJ$, each block $(x,y)$ has projections equal to those of one of
$D^c$, for some $c$ (Requirement~2).  We show the construction of such
block packings in the next section.


\section{Construction of Block Packings}\label{sec: construction of block packings}

As in the previous section, $T$ is a fixed (but arbitrary) non-bar
tile.  We call $(i,j)$ a \emph{conflicting} vector if $T$ and
$T+(i,j)$ overlap, that is $T\cap (T+(i,j))\neq\emptyset$. Obviously,
the vectors $(i,j)$ and $(-i,-j)$ are either both conflicting or both
non-conflicting. Since $T$ is not a bar, it has a conflicting vector
$(i, j)$ with $i, j \neq 0$.  To see this, observe that since $T$ is connected
and not a bar, it contains two cells $(i,j)$ and $(i',j')$ with
$|i-i'|=1$ and $|j-j'|=1$, so one of $(1, 1)$ or $(-1, 1)$
is a conflicting vector.

For the construction of the proof, fix a conflicting vector of $T$
that maximizes the $L_1$ norm under the constraint that none of the
coordinates is $0$. We denote this vector by $(-p,q)$ and assume
without loss of generality $p,q > 0$, for otherwise we can flip $T$
horizontally or vertically and give the proof for the resulting tile.

So any vector $(i, j)$ with $i,j \neq 0$ and $|i |
+ |j | > |p| + |q|$ is not conflicting.  
Let $a$ be the smallest positive integer such that $(a p, 0)$ is not a
conflicting vector. Similarly let $b$ be the smallest positive integer
such that $(0, b q)$ is not a conflicting vector.  Without loss of
generality we assume that $a \leqslant b$, since otherwise we can
exchange the roles of columns and rows in the proof.

We now divide the proof into four cases, and for each of them we show
that Requirements 1 and 2 are satisfied.

\begin{figure}[h]
\centerline{\includegraphics[width=11.75cm]{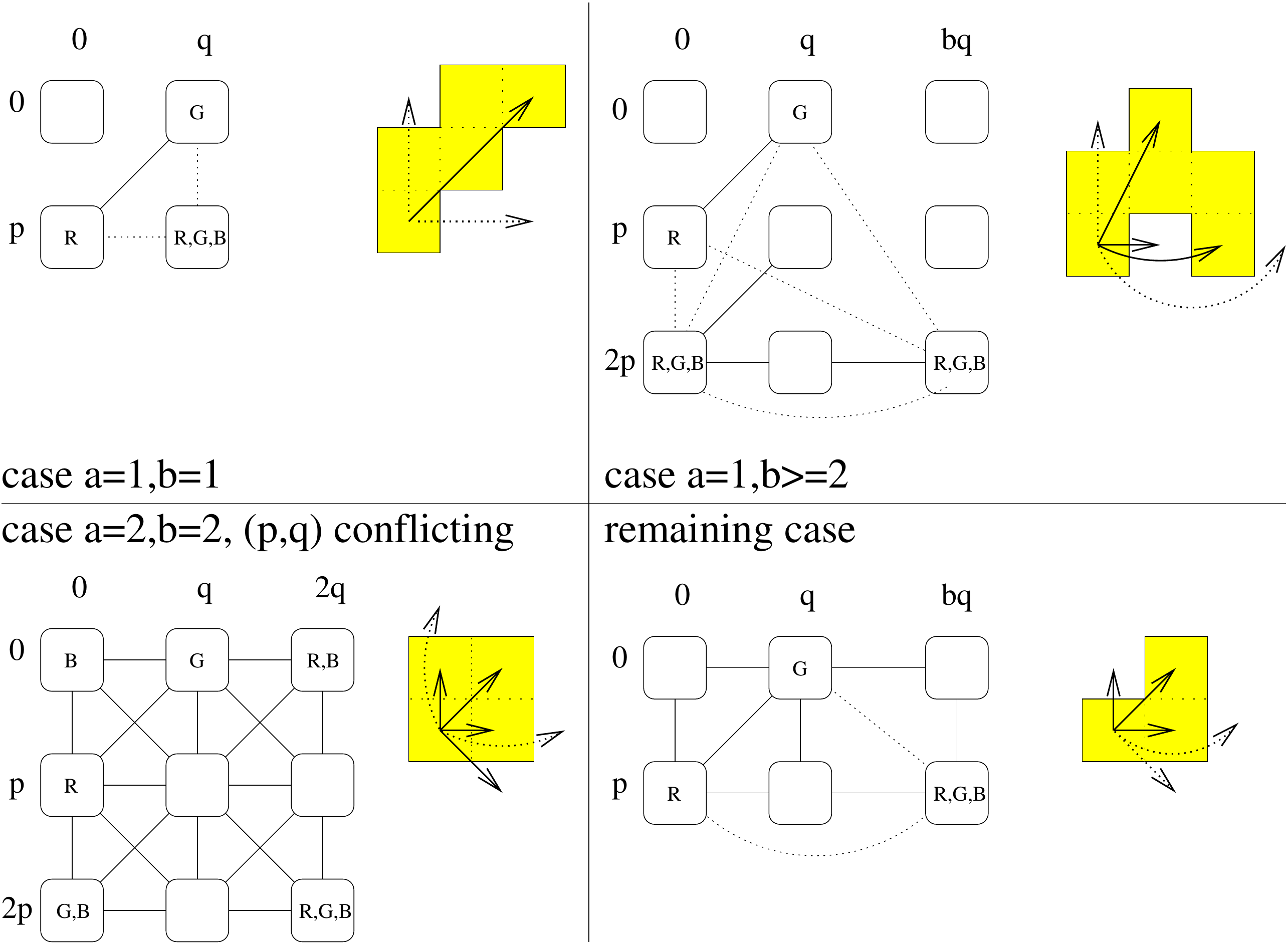}}
\caption{For each case of the proof, the figure shows
three packings (left-hand side)
and an example of a tile satisfying this case (right-hand side). 
Dotted vectors are non-conflicting, either by
maximality of $(-p,q)$ or by the case condition. In the third case,  all vectors 
	that are not shown (to avoid clutter) are non-conflicting.}
	\label{fig:a1b1} 
	\label{fig:a1b2} 
	\label{fig:a2b2with} 
	\label{fig:a2b2without} 
\end{figure}


\subsection{Case $a = 1$ and $b = 1$}

In this case, we use the following three packings:
\begin{align*}
 D^{R} &= \braced{(p, 0), (p, q)},
	\\
 D^{G} &= \braced{(0, q), (p, q)},
	\\
	 D^{B} &=\braced{(p, q)}.
\end{align*}
The values of $k$ and $\ell$ are chosen to be the smallest integers for which
these three packings are contained in the $k\times\ell$ grid.


The packings are depicted on Figure~\ref{fig:a1b1}. The squares represent
possible positions for tiles. Two positions are connected with a solid line if
the difference of the positions is a conflicting vector. That means that no
packing can contain simultaneously a tile in both positions.
Dotted lines indicate non-conflicting vectors, i.e. they connect pairs
of compatible positions. We show lines only between position pairs relevant to 
the proof. In the figure we mark with letter $c$ the tile positions of $D^c$, for
$c\in\braced{R,G,B}$. For illustration, for each case, on the right-hand side
of the figure we show a tile satisfying the case conditions. 
Again, solid vectors are conflicting and dotted vectors are 
non-conflicting.

We first verify Requirement~1. The $(0,p)$-restrictions of 
$\barr^R, \barr^G, \barr^B$ are, respectively, $(0,2)$, $(1,1)$ and $(0,1)$, and the $(0,q)$-restrictions of
$\bars^R, \bars^G, \bars^B$ are, respectively, $(1,1)$, $(0,2)$ and $(0,1)$.
For both the row and column projections, routine calculations show that their
restrictions are affinely independent.

We now focus on Requirement~2.
Let $r,s$ be the projections obtained by the reduction, and consider a packing $D$ 
with these projections. We use Lemma~\ref{lem:ryser-adapted}, with
$I$ being the set of all row indices that are $p$ modulo $k$, and $J$ being
the set of all column indices that are $q$ modulo $\ell$. 
By inspecting the definition of the projections we have
\[
r(I) = \sum_{x\in[0,m)} r^R_x + mn = \sum_{y\in[0,n)} s^R_y + mn =
s(\barJ)+mn,
\] 
so $r(I)-s(\barJ)=mn$, 
which is $|I\times J|$. In consequence Lemma~\ref{lem:ryser-adapted} applies.
Therefore every block in $D$ contains a tile at position $(p,q)$ and none at 
position $(0,0)$.  The remaining possible positions for tiles are 
$(p,0),(0,q)$, but both cannot be occupied in the same block. 
This shows that every block of $D$ is one of the packings $D^{R},D^{G},D^{B}$.


\subsection{Case $a = 1$ and $b \geq 2$}

In this case, we use the following three packings:
\begin{align*}
D^{R} &= \braced{(2p,0), (2p,bq) ,(p,0)} 
\\
D^{G} &= \braced{(2p,0), (2p,bq) , (0,q)}
\\
D^{B} &= \braced{(2p,0), (2p,bq)} 
\end{align*}
The values of $k$ and $\ell$ are chosen to be the smallest integers for which
these three packings are contained in the $k\times\ell$ grid. 

The $(0,p,2p)$-restrictions of $\barr^R, \barr^G, \barr^B$ are
linearly independent vectors $(0,1,2)$, $(1,0,2)$, $(0,0,2)$;
therefore $\barr^R, \barr^G, \barr^B$ are affinely independent. By a
similar argument, we obtain that the corresponding column projection
vectors $\bars^R, \bars^G, \bars^B$ are affinely independent as well.
Thus, Requirement~1 holds.

Now we verify Requirement~2.  Row $2 p$ of a block can contain at
most 2 tiles.  Let $I$ be the set of rows $i$ with $i\bmod k=2p$.  By
inspecting the definition of the projections we have $r(I)=2mn$, so
every block in a solution $D$ must contain exactly 2 tiles in row
$2p$, and they are at positions $(2p,0),(2p,bq)$.  

Now let $J$ be the
set of all columns $j$ with $j\bmod \ell=bq$. We only have $s(J)=mn$,
so in every block of $D$ the positions $(0,bq),(p,bq)$ are empty.
The tile at $(2p,0)$ forces position $(p, q)$ to be empty.  By the
case assumption that $a=1$, there is no conflict between $(2p, 0)$ and $(p,
0)$.  By maximality of $(-p,q)$ there is no conflict between positions
$(2 p, 0)$ and $(0, q)$, or between $(2 p, b q)$ and $(p, 0)$, or $(2 p,
b q)$ and $(0, q)$.  

That leaves only $3$ positions where the block
packings can differ, namely $(0,0)$, $(p,0)$ and $(0,q)$.  
Let $d$ be the number of blocks in $D$ with a
tile in $(0, 0)$.  Similarly let $e$ be the number of blocks in $D$
with a tile in $(0,q)$.  Now we use the fact that in the original
instance we had $\sum r^{G}_{x} = \sum s^{G}_{y}$; let $\lambda_G$ denote
this quantity.  This time, let $I'$ be the set of rows $i$ with $i\bmod
k=0$, and $J'$ the set of columns $j$ with $j\bmod \ell=q$.  Then,
by the definition of $d$ and $e$, we have $r(I')=d+e$,
and by the definition of the chosen three packings, we have $r(I')=\lambda_G$. 
Similarly, we have $r(J')=e$ and $r(J')=\lambda_G$,
which shows $d=0$.  Therefore every block packing in $D$ is one
of $D^{R},D^{G}$ or $D^{B}$.


\subsection{Case $a = 2$, $b = 2$ and Vector $(p, q)$ Conflicting}

In this case we assume $a = 2$, $b = 2$ and that the vector $(p, q)$
is conflicting. Since $(-p,q)$ is conflicting as well, this makes the
construction very symmetric. The three packings used in this
case are:
\begin{align*}
 D^{R} &=\braced{(0,2q), (p,0), (2p, 2q)},
 \\
 D^{G} &= \braced{(0,q), (2p,0) , (2p, 2q)},
 \\
 D^{B} &= \braced{(0,0), (0,2q), (2p,0), (2p,2q)}.
\end{align*}
Again, the values of $k$ and $\ell$ are chosen to be the smallest
integers for which these three packings are contained in the
$k\times\ell$ grid.  The construction is illustrated in
Figure~\ref{fig:a2b2with}. The idea behind this construction is
similar to the reduction used in \cite{CCDW01} to show {\NP}-hardness
of the packing problem for the $2 \times 2$ square tile.

The $(0,p,2p)$-restrictions of $\barr^R, \barr^G, \barr^B$ are
$(1,1,1)$, $(1,0,2)$, $(2,0,2)$; therefore $\barr^R, \barr^G, \barr^B$
are affinely independent. By symmetry, the same holds for $\bars^R,
\bars^G, \bars^B$.


Now we verify Requirement~2. Let $D$ be a packing with projections
$r,s$.  Due to conflicts, there can be at most 2 tiles in
rows $p,2p$ of a block. Let $I=\{i:i\bmod k\in\{p,2p\}\}$. Since
$r(I)=2mn$ and there are $mn$ blocks in $D$, 
every block of $D$ must
contain exactly 2 tiles in rows $p,2p$. By symmetry, the same holds for
columns $q,2q$. There are only 4 packings that satisfy these constraints
and avoid conflicts, namely $D^R,D^G,D^B$ and packing
$D^A=\{(0,2q),(2p,0),(2p,2q)\}$. Let $\lambda_R,\lambda_G,\lambda_B,\lambda_A$ be the
respective numbers of these different block packings in $D$. 
Since $D$ has row projection $r$, by expressing the total number of tiles in
two different ways, we have
\begin{align}
	3\sum r_x^R + 3\sum r_x^G + 4\sum r_x^B  
			= 3\lambda_R+ 3\lambda_G + 4\lambda_B + 3\lambda_A.
						\label{eqn: case 3-1}
\end{align}
From the assumption that $\sum_c \sum_x r^c_x = mn$, we have also
\begin{align}
	\sum r_x^R + \sum r_x^G + \sum r_x^B = mn = \lambda_R+ \lambda_G 
						+ \lambda_B+\lambda_A.
		\label{eqn: case 3-2}
\end{align}
Now let $I'=\{i: i\bmod k=p\}$. Then $\sum_x r^R_x = r(I') =\lambda_R$.
Similarly, for $J'=\{j:j\bmod \ell=q\}$ we obtain
$\sum_x r^G_x = \sum_y s_y^G = s(J') =\lambda_G$. 
After subtracting these two equations from (\ref{eqn: case 3-1})
and (\ref{eqn: case 3-2}), we are left with two equations
\begin{align*}
	\sum r_x^B  &= \lambda_B + \lambda_A, &
	 4\sum r_x^B &= 4\lambda_B + 3\lambda_A,
\end{align*}
from which we conclude $\lambda_A=0$. This verifies Requirement~2.


\subsection{The Remaining Case}

Assume now that none of the previous cases holds. Since
$a\le b$, this means that either
\begin{description}
	\item{(a)} $a \geq 2$ and $b \geq 3,$ or
	\item{(b)} $a = 2$, $b = 2$ and vector $(p, q)$ is not conflicting.
\end{description}
We claim that the vector $(p,(b-1)q)$ is not
conflicting. Indeed, in case (b) above, it follows by case assumption
and in case (a), it follows from the maximality of $(-p,q)$. Therefore,
in any block of a $T$-packing both positions $(0,q)$ and $(p,bq)$ could contain
a tile.

We use the following three packings (see Figure~\ref{fig:a2b2without}):
\begin{align*}
 D^{R} &= \braced{(p,0), (p, bq)},
	\\
 D^{G} &= \braced{(0,q), (p,bq)},
	\\
	 D^{B} &= \braced{(p,bq)}.
\end{align*}
Again, the values of $k$ and $\ell$ are chosen to be the smallest
integers for which these three packings are contained in the
$k\times\ell$ grid.

The $(0,p)$-restrictions of $\barr^R, \barr^G, \barr^B$ are $(0,2)$,
$(1,1)$, $(0,1)$, and their difference vectors $(1,-1)$, $(0,-1)$ are
linearly independent. Therefore $\barr^R, \barr^G, \barr^B$ are affinely
independent.  The $(0,q,bq)$-restrictions of $\bars^R, \bars^G,
\bars^B$ are linearly independent vectors $(1,0,1)$, $(0,1,1)$,
$(0,0,1)$; therefore $\bars^R, \bars^G, \bars^B$ are affinely
independent. Thus, Requirement~1 holds.


Now we verify Requirement~2. The proof is similar to those of
previous cases, except that now we have more candidate packings
to consider. Fix some $T$-packing $D$ with projections
$r,s$.  First, using the same arguments as in the previous case,
we observe that every block of $D$ must contain exactly
one tile in column $bq$, that is either in location $(0,bq)$ 
or $(p,bq)$. Taking conflicts into accounts, straightforward
case analysis produces nine possible packings, including two
pairs of ``equivalent" packings with identical projections.
We now introduce notation for the numbers of these packings:
\begin{description}
  \item{$\lambda_R =$} the number of packings $D^R = \braced{(p, 0), (p, bq)}$,

  \item{$\lambda_G =$} the number of packings  $D^G= \braced{ (0, q), (p, bq)}$ 
                or $\{(p,q),(0,bq)\}$,
  
  \item{$\lambda_B =$} the number of packings $D^B = \braced{(p,bq})$,
  
  \item{$\lambda_A =$} the number of packings $\braced{ (0,0), (p,bq)}$
			or $\braced{(p, 0), (0, bq)}$,
  
  \item{$\lambda_C=$} the number of packings $\braced{(0, 0), (p, q), (0,bq)}$,
  
  \item{$\lambda_D =$} the number of packings $\braced{(0, bq)}$,

  \item{$\lambda_E=$} the number of packings $\braced{(0,0), (0, bq)}$.
\end{description}

Let $I$ be the set of all rows $i$ with $i\bmod k=0$ and $J$ be the set
of all columns $j$ with $j \bmod \ell=q$. Then, by inspecting the
projection definitions, we have 
$r(I)=\sum_{x} r^{G}_{x} =\sum_{y} s^{G}_{y}=s(J)$. 
Since $r$ and $s$ are the projections
of $D$, we also have 
$r(I)= \lambda_G + \lambda_A + 2 \lambda_C + \lambda_D + 2\lambda_E$ 
and $s(J)=\lambda_{G}+\lambda_{C}$.  This shows $\lambda_A = \lambda_C = \lambda_D
= \lambda_E = 0$,
completing the analysis of this case and the proof of {\NP}-hardness.


\section{Acknowledgements}

This research was partially supported by the USA National Science Foundation, grant CCF-0729071, 
and by the Spanish CICYT ({\em Comisi\'on interministerial de ciencia y tecnolog\'{\i}a}) 
projects TIN2007-68005-C04-03 and TIN2008-06582-C03-01.

We would like to thank the anonymous reviewers for pointing out a number of typographical
mistakes in the submitted manuscript.


\small
\bibliographystyle{plain} 
\bibliography{tiling}

\end{document}

%% file: macros.tex
\newcommand{\barr}{{\bar r}}
\newcommand{\bars}{{\bar s}}

\newcommand{\barI}{{\bar I}}
\newcommand{\barJ}{{\bar J}}

\newcommand{\calI}{{\cal I}}
\newcommand{\calJ}{{\cal J}}

\newcommand{\braced}[1]{{ \left\{ #1 \right\} }}

\newcommand{\parend}[1]{{ \left( #1 \right) }}
\newcommand{\barred}[1]{{ \left| #1 \right| }}

\newcommand{\suchthat}{{\,:\,}}

\newcommand{\reals}{{\mathbb R}}

\newcommand{\NP}{{\mbox{$\mathbb{NP}$}}}

\newcommand{\TPTP}[1]{{\mbox{\sf TPTP$(#1)$}}}
\newcommand{\threeCTP}{{\mbox{\sf 3CTP}}}

\newenvironment{myproof}{{\it Proof:\/}}{\hspace*{1em}\hfill$\Box$\vskip 0.1in}

\newcommand{\ignore}[1]{}